\documentclass[twocolumn,preprintnumbers,showpacs,amsmath,amssymb,floatfix]{revtex4}

\usepackage{amssymb}
\usepackage{mathrsfs}

\usepackage{graphicx}
\usepackage{dcolumn}
\usepackage{bm}
\usepackage{mathrsfs}
\usepackage{multirow}   
\usepackage{hyperref}

\usepackage{atlasphysics}

\usepackage{lldefs}

\newcommand{\LumiElectronsInFb}{1.08}
\newcommand{\LumiMuonsInFb}{1.21}

\newcommand{\LimitElectronExpected}{1.70}
\newcommand{\LimitElectron}{1.70}

\newcommand{\LimitMuonExpected}{1.61}
\newcommand{\LimitMuon}{1.61}

\newcommand{\LimitCombinedExpected}{1.83}
\newcommand{\LimitCombined}{1.83}

\newcommand{\LimitElectronExpectedG}{1.50}
\newcommand{\LimitElectronG}{1.51}

\newcommand{\LimitMuonExpectedG}{1.44}
\newcommand{\LimitMuonG}{1.45}

\newcommand{\LimitCombinedGOne}{0.71}
\newcommand{\LimitCombinedGThree}{1.03}
\newcommand{\LimitCombinedGFive}{1.33}
\newcommand{\LimitCombinedExpectedG}{1.63}
\newcommand{\LimitCombinedG}{1.63}

\newcommand{\LimitCombinedSq}{1.60}
\newcommand{\LimitCombinedN}{1.52}
\newcommand{\LimitCombinedPsi}{1.49}
\newcommand{\LimitCombinedChi}{1.64}
\newcommand{\LimitCombinedEta}{1.54}
\newcommand{\LimitCombinedI}{1.56}

\usepackage{epstopdf}
\usepackage{multirow}
\usepackage[hang,center]{subfigure}

\begin{document}

\title{Search for dilepton resonances in $pp$ collisions at $\sqrt{s}=7$~TeV with the ATLAS detector} 
\begin{minipage}[t]{\textwidth}
  \begin{flushright}
    CERN-PH-EP-2011-123\\
    Submitted to Physical Review Letters
    \end{flushright}
\end{minipage}
\vspace{0.5cm}
      
\author{The ATLAS Collaboration}

\begin{abstract}
This Letter reports on a search for narrow high-mass resonances decaying into dilepton final states.
The data were recorded by the ATLAS experiment in $pp$ collisions at $\rts = 7$~\TeV\ at the Large Hadron Collider (LHC)  
and correspond to a total integrated luminosity of \LumiElectronsInFb\  (\LumiMuonsInFb)~\ifb\ in the \ee\  (\mumu) channel. 
No statistically significant excess above the Standard Model expectation is observed 
and upper limits are set at the 95\% C.L. on the cross section times branching fraction of \zp\ resonances and Randall-Sundrum gravitons decaying into 
dileptons as a function of the resonance mass. 
A lower mass limit of \LimitCombined~TeV on the Sequential Standard Model \zp\ boson is set. 
A Randall-Sundrum graviton with coupling $k/\overline{M}_{Pl}$=0.1 is excluded at 95\%~C.L. for masses below \LimitCombinedG~TeV.
\end{abstract}


\maketitle

This Letter describes a search for narrow high-mass resonances decaying into \ee\ or \mumu\ pairs   
using 7~\TeV\ $pp$ collision data recorded 
with the ATLAS detector~\cite{atlas:detector}.
Such resonances, which  are predicted by several extensions of the Standard Model (SM), include 
new  heavy spin-1 neutral gauge bosons such as \zp~\cite{London:1986dk,Langacker:2008yv,Erler:2009jh} and 
\zstar~\cite{wzstar}, techni-mesons~\cite{Lane:1989ej,Lane:2003,Belyaev:2009}, 
as well as  spin-2  Randall-Sundrum gravitons, $G^*$~\cite{RS}. 

The benchmark models considered for the \zp\ are the
Sequential Standard Model (SSM)~\cite{Langacker:2008yv}, with the same couplings to fermions as the \z\ boson, 
and the \esix\ grand unified symmetry group~\cite{London:1986dk}, broken into \su5\ and two
additional \u1\ groups, leading to new neutral gauge fields $\psi$ and $\chi$.
The particles associated with the additional fields 
can mix in a linear combination to form the \zp\ candidate: 
$\zp (\te6 ) = \zp _\psi \cos{\te6 } + \zp _\chi \sin{\te6 }$, 
where \te6 is the mixing angle between the two gauge bosons.   
The pattern of spontaneous symmetry breaking and the value of \te6\ determine the \zp\ couplings to fermions;
six well-motivated choices of \te6~\cite{Langacker:2008yv,London:1986dk} lead to the specific \zp\ states 
named \zppsi, \zpN, \zpeta, \zpI, \zpsq\ and \zpchi.

Other models predict additional spatial dimensions as a possible explanation for the gap between the electroweak 
symmetry breaking scale and the gravitational energy scale. 
The Randall-Sundrum (RS) model~\cite{RS}  
predicts excited Kaluza-Klein modes of the graviton, which appear as spin-2 resonances. 
These modes have a narrow intrinsic width when $k/\overline{M}_{Pl}<$ 0.1, 
where $k$ is the spacetime curvature in the extra dimension,  and
$\overline{M}_{Pl}  = M_{Pl}/\sqrt{8\pi}$ is the
reduced Planck scale.

Previous searches
 have set direct and indirect     constraints on the mass of the $G^*$ and  \zp\ resonances~\cite{resonanceReview,Langacker:2009su}.
The Tevatron~\cite{Abazov:2010ti,CDF:Zpmumu} experiments exclude a \zpssm\ with a mass lower than  
1.071~TeV~\cite{CDF:Zpmumu}. Recent measurements from the LHC experiments, based on $\approx$40~pb$^{-1}$ of data recorded in 2010,  
exclude a \zpssm\ with a mass lower than  1.042~TeV (ATLAS)~\cite{Aad:2011xp} and  1.140~TeV (CMS)~\cite{CMS_Zp2010}.  
Indirect constraints from LEP~\cite{Abbiendi:2003dh,Abdallah:2005ph,Achard:2005nb,Schael:2006wu}  
  extend these limits to 1.787~TeV~\cite{Langacker:2009su}.
Constraints on the mass of the RS graviton have been set by the CMS~\cite{CMS_Zp2010}, CDF~\cite{Aaltonen:2011xp} and D0~\cite{Abazov:2010xh} 
collaborations, excluding RS gravitons with mass below 1.079~TeV for $k/\overline{M}_{Pl}=$ 0.1~\cite{CMS_Zp2010}.

The ATLAS detector 
consists of inner tracking devices surrounded by a 2~T superconducting solenoid, 
electromagnetic and hadronic calorimeters, and a muon spectrometer with a toroidal magnetic field. 
Charged particles 
in the pseudorapidity range $|\eta| < 2.5$~\cite{coordinates} 
are reconstructed with the inner detector, which consists of silicon pixel, silicon strip, and transition radiation detectors.
The superconducting solenoid is surrounded by a hermetic calorimeter that covers $|\eta| < 4.9$.
For $|\eta| < 2.5$, the electromagnetic calorimeter is finely segmented and plays an important role in electron identification.
Outside the calorimeter, air-core toroids provide the magnetic field for the muon spectrometer. 
Three sets of precision drift tubes and cathode strip chambers 
provide an accurate measurement of the muon track curvature in the region $|\eta |< 2.7$.
Resistive-plate and thin-gap chambers provide muon triggering capability up to $|\eta |< 2.4$.

The data sample used in this analysis, recorded during the first half of 2011, corresponds to 
a total integrated luminosity of \LumiElectronsInFb\  (\LumiMuonsInFb)~\ifb\ in the \ee\  (\mumu) channel.
Events are required to pass single electron (muon) triggers with a transverse energy \et\ (transverse momentum \pt) threshold above 20 (22)~\gev.
Collision candidates are selected by requiring a primary vertex  with at least three associated 
charged particle tracks with $\pt > 0.4~\gev$.

In the \ee\ channel,
two electron candidates are required with transverse energy $E_{T} > 25$~\gev\ and 
$|\eta| < 2.47$; 
the transition region $1.37\leq |\eta|\leq 1.52$ between the barrel and the endcap calorimeters is excluded.
Electron candidates are 
formed from clusters of cells reconstructed in the electromagnetic calorimeter associated with a charged particle track in the inner detector. 
Criteria on the transverse shower shape, the longitudinal leakage into the hadronic calorimeter,
and the association with an inner detector track are applied to the cluster to define a so-called \Medium\ electron~\cite{atlas:egamma_id_900GeV,Aad:2010yt}.
The electron energy is obtained from the calorimeter measurement and its direction from the associated track.
A hit in the first active pixel layer is required to suppress background from photon conversions.
To further suppress background from QCD jet production, the higher $E_{T}$ electron is required to be isolated by demanding that 
 $\Sigma E_T(\Delta R<0.2)<7\gev$, where $\Delta R=\sqrt{(\Delta \eta)^2 + (\Delta \phi)^2}$ 
and  $\Sigma E_T (\Delta R < 0.2)$ is the sum of the transverse energies around the electron direction. The core of the electron energy deposition is excluded and
the sum is corrected for transverse shower leakage and pile-up from additional $pp$ collisions. 
The two electron candidates are not required to have opposite charge to minimize the impact of possible charge mis-identification.
For these selection criteria, the total signal acceptance for a \zpee\ (\gee) of mass 1.5~\tev\  is 65\% (69\%), and is approximately independent 
 of mass above 600 GeV.
These numbers include the acceptance of all selection cuts and  efficiencies, and reflect the lepton angular
 distributions due to spin. 

In the \mumu\ channel,
two muon candidates of opposite charge are required, each satisfying $\pt > 25$~\gev.
Muon tracks are reconstructed independently in both the inner detector and muon spectrometer, and  
their momenta are determined from a combined fit to these two measurements. 
To optimize the momentum resolution, 
each muon candidate is required to pass quality cuts in the inner detector 
and to have at least three hits in each of the inner, middle, and outer layers of the muon system. 
Muons with hits in both the barrel and the endcap regions are discarded because 
of residual misalignment between these two parts of the muon spectrometer. 
The effects of misalignments and intrinsic position resolution are included in the simulation. The \pt\ resolution
at 1~TeV ranges from 15\% (central) to 44\% (for $|\eta |> 2$).

To suppress background from cosmic rays, the muon tracks are  required to have a transverse impact parameter $|d_0|<0.2$~mm, 
a distance along the beam-line to the primary vertex below 1~mm, 
and the $z$ position of the primary vertex $|z({\rm PV})|<200$~mm.
To reduce background from QCD jets, each muon is required to be isolated such that 
 $\Sigma\pt(\Delta R<0.3)/\pt(\mu)<0.05$, where only tracks with $\pt >1$~GeV enter the sum. 
  The total signal acceptance 
  is 40\% (44\%) for a \zpmumu\ (\gmumu) of mass 1.5~\tev. 
The lower acceptance compared to the electron channel is 
due to the stringent 
requirements on the muon selection criteria 
to improve $p_T$ resolution.

For both channels, the dominant and irreducible background is due to 
the $Z/\gamma^*$ (Drell-Yan) process, characterized by the same final state as the signal.  
Small contributions from \ttbar\ and diboson ($WW$, $WZ$ and $ZZ$) production are also 
present in both channels. Semi-leptonic decays of $b$ and $c$ quarks in the \mumu\ sample
and a mixture of photon conversions, semi-leptonic heavy quark decays, and hadrons faking electrons in the \ee\ sample
are backgrounds that are referred to below as QCD background.
Jets accompanying $W$ bosons (\wpjet) may similarly produce lepton candidates.  

The expected signal and backgrounds, with the exception of the QCD component,
are evaluated with simulated 
samples and rescaled using the most precise available cross section predictions.
The \zp, \gstar\ signal and $Z/\gamma^*$ processes are generated with
\pythia\ 6.421~\cite{Sjostrand:2006za} using MRST2007 LO*~\cite{mrst} parton distribution functions (PDFs).
Interference between the $Z/\gamma^*$ processes and the heavy resonances is small and therefore neglected.
The diboson processes are generated with \herwig~6.510~\cite{herwig} using MRST2007 LO* PDFs.
The \wpjet\ background is generated with \alpgen~\cite{Mangano:2002ea} using CTEQ6L1~\cite{Pumplin:2002vw} PDFs and
the \ttbar\ background with \mcatnlo~3.41~\cite{mcatnlo} using CTEQ66~\cite{Nadolsky:2008} PDFs.
For both, \jimmy~4.31~\cite{jimmy} is used to describe multiple parton interactions and
\herwig\ to describe the remaining underlying event and parton showers.
Final-state photon radiation is handled with \photos~\cite{fsr_ref}. 
The samples are processed through a full ATLAS detector simulation~\cite{atlas:sim}
   based on GEANT4~\cite{geant}.

The $Z/\gamma^*$ cross section is calculated at next-to-next-to-leading order (NNLO)
using PHOZPR~\cite{Hamberg:1990np} with MSTW2008 PDFs~\cite{mstw}. 
The ratio of this cross section to the leading-order cross section is used to determine a mass dependent QCD K-factor which is 
applied to the results of the leading-order simulations. 
The same QCD K-factor is applied to the \zp\ signal. No 
QCD K-factor is available for  \gstar\ production at 7 TeV~\cite{Kumar:2009epj,Matthews:2009np}.  
 Higher-order weak corrections (beyond the photon radiation included in the simulation) 
are calculated using \horace~\cite{horace,CarloniCalame:2007cd}, yielding a weak K-factor due to virtual heavy gauge boson loops.
The weak K-factor is only applied to the Drell-Yan background.
The diboson cross sections are calculated to next-to-leading order (NLO) using {\sc mcfm}~\cite{Campbell:1999mcfm} with an uncertainty of 5\%.
The \wpjet\ cross section is 
rescaled to the inclusive NNLO calculation of {\sc fewz}~\cite{Melnikov:2006kv},
resulting in 30\% uncertainty when at least one parton with $E_{T} > 20$~\gev\ accompanies the \w\ boson.
The \ttbar\ cross section is  predicted at approximate-NNLO, with 10\% uncertainty~\cite{Moch:2008qy,Langenfeld:2009tc}.

The QCD background in the \ee\ sample is estimated with data using  ``reversed electron identification'' and ``isolation fit'' techniques~\cite{Aad:2011xp}, and a third method that uses fake rates measured from inclusive jet samples.
In the ``reversed electron identification'' technique, data with 
both electron candidates failing some  identification criteria (chosen not to affect kinematic distributions) 
are used to determine the 
 QCD background distribution versus $m_{ee}$. This method is used for the central estimate and the others which bracket it,
 to assign a systematic uncertainty.
  The QCD background in the \mumu\ sample 
is evaluated from data using the muon isolation variable $\Sigma \pt (\Delta R < 0.3) / \pt$
\cite{Aad:2011xp}. 
The QCD and \wpjet\ backgrounds are small (negligible) for the electron (muon) channel.
 Backgrounds from cosmic rays 
are 
negligible.

The observed invariant  mass distributions are compared to the 
 SM expectation. For this purpose, 
 the Drell-Yan, $t \bar{t}$, diboson and \wpjet\ backgrounds from Monte Carlo simulation are 
 scaled according to their respective cross sections and added to the  QCD background. 
The simulated backgrounds are  then rescaled so that the sum matches the observed number of data events
in the 70 - 110 GeV mass interval. The scaling factor is within 1\% of unity. 
The advantage of this approach is 
that the uncertainty on the luminosity, and any mass independent uncertainties on efficiencies, cancel between the \zp\ (\gstar) and the \z\ boson.

Figure~\ref{fig:mll} presents the invariant mass ($m_{\ell\ell}$) distribution for the dielectron (top) and dimuon (bottom) final states after final selection,
while 
Table~\ref{tab:backgroundTableCombo}  
shows the number of data events and the estimated backgrounds in bins of 
reconstructed $m_{\ell\ell}$. 
The dilepton invariant mass distributions are well described by the prediction from SM processes.
Figure~\ref{fig:mll} also displays the expected \zpssm\ signal for three mass hypotheses. 

\begin{figure}[!tb]
  \includegraphics[width=\columnwidth]{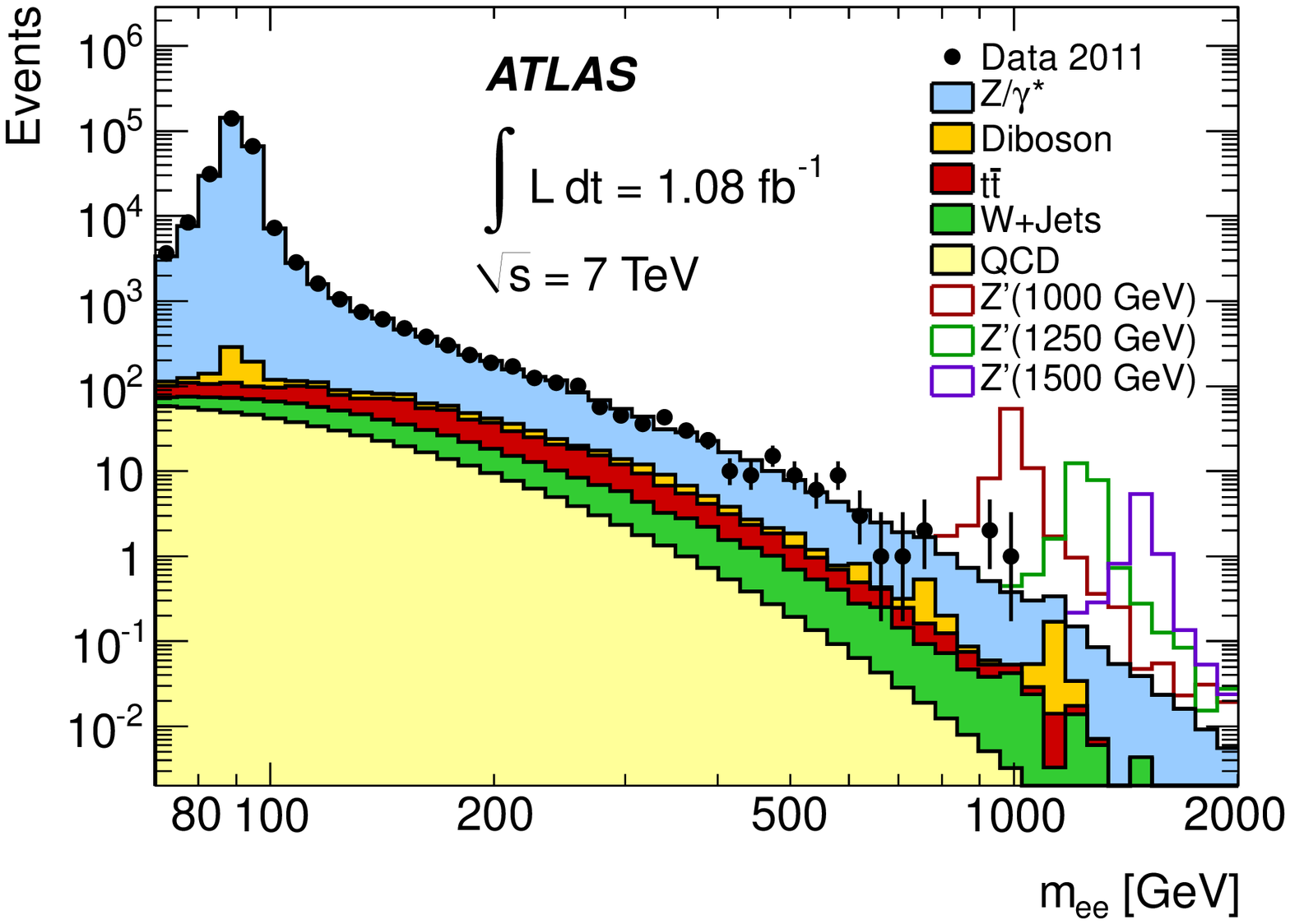}
  \includegraphics[width=\columnwidth]{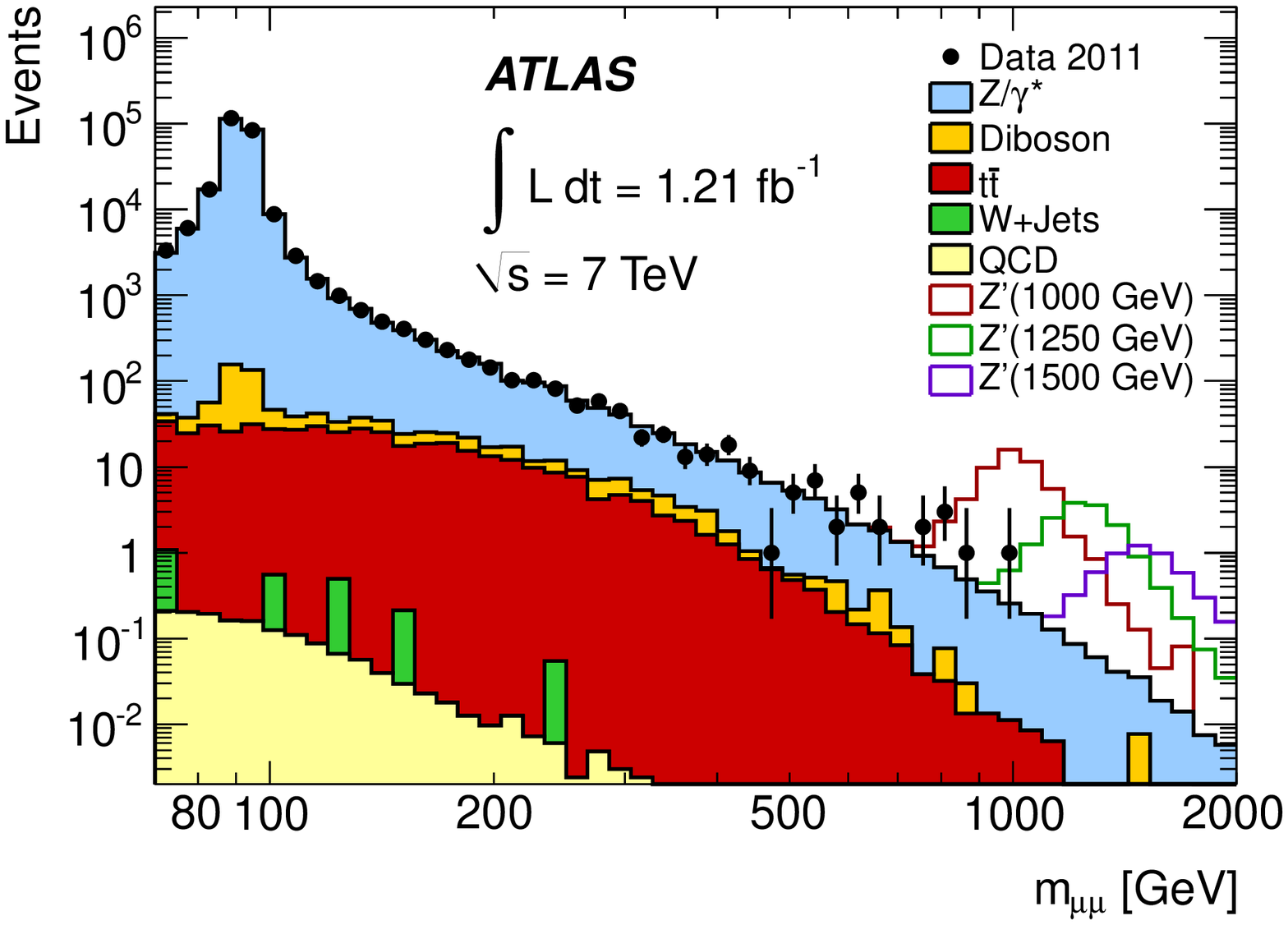}  
  \caption{Dielectron (top) and dimuon (bottom) invariant mass ($m_{\ell\ell}$) distribution after final selection, compared to the stacked sum of all expected backgrounds,
  with three example \zpssm\ signals overlaid. The bin width is constant in $\log m_{\ell\ell}$.}
  \label{fig:mll}
\end{figure}

\begin{table}[!htb]
\caption{Expected and observed number of events in the dielectron (top) and dimuon (bottom) channels. 
The first bin is used to normalize the total background to the data. 
The errors quoted include both statistical and systematic uncertainties, except the error on the total background in the normalization region 
which is given by the square root of the number of observed events. 
The systematic uncertainties are correlated across bins and are discussed in the text.  
}
\vspace{5mm}
\centering
\small\addtolength{\tabcolsep}{+2pt}
\resizebox{\columnwidth}{!}{
\begin{tabular}{cccccc}
\hline
\mee [GeV]& 70-110& 110-200  & 200-400 & 400-800 & 800-3000     \\
\hline
DY & 258482 $\pm$ 410 & 5449 $\pm$ 180 & 613 $\pm$ 26 & 53.8 $\pm$ 3.1 & 2.8 $\pm$ 0.1 \\
\ttbar & 218 $\pm$ 36 & 253 $\pm$ 10 & 82 $\pm$ 3 & 5.4 $\pm$ 0.3 & 0.1 $\pm$ 0.0 \\
Diboson & 368 $\pm$ 19 & 85 $\pm$ 5 & 29 $\pm$ 2 & 3.1 $\pm$ 0.5 & 0.3 $\pm$ 0.1 \\
W+jets & 150 $\pm$ 100 & 150 $\pm$ 26 & 43 $\pm$ 10 & 4.6 $\pm$ 1.8 & 0.2 $\pm$ 0.4 \\
QCD & 332 $\pm$ 59 & 191 $\pm$ 75 & 36 $\pm$ 29 & 1.8 $\pm$ 1.4 & $< 0.05$ \\
\hline
 Total & 259550 $\pm$ 510 & 6128 $\pm$ 200 & 803 $\pm$ 40 & 68.8 $\pm$ 3.9 & 3.4 $\pm$ 0.4 \\
\hline
 Data & 259550 & 6117 & 808 & 65 & 3 \\
\hline
\\

\\ 

\hline
$m_{\mu^{+}\mu^{-}}$ [GeV]  &  70-110  &  110-200  &  200-400  &  400-800  &  800-3000  \\
\hline
DY       &  236319 $\pm$ 320  &  5171 $\pm$ 150  &  483 $\pm$ 22  &  40.3 $\pm$ 2.5  &  2.0 $\pm$ 0.3  \\ 
\ttbar    &  193 $\pm$ 21  &  193 $\pm$ 20  &  63 $\pm$ 6  &  4.2 $\pm$ 0.4  &  0.1 $\pm$ 0.0  \\ 
Diboson  &  307 $\pm$ 16  &  69 $\pm$ 5  &  25 $\pm$ 2  &  1.7 $\pm$ 0.5  &  $< 0.05$  \\ 
\w+jets   &  1 $\pm$ 1  &  1 $\pm$ 1  &  $< 0.5$  &  $< 0.05$  &  $< 0.05$  \\ 
QCD      &  1 $\pm$ 1  &  $< 0.5$  &  $< 0.5$  &  $< 0.05$  &  $< 0.05$  \\ 
\hline
Total    &  236821 $\pm$ 487  &  5434 $\pm$ 150  &  571 $\pm$ 23  &  46.1 $\pm$ 2.6  &  2.1 $\pm$ 0.3  \\ 
\hline
Data     &  236821  &  5406  &  557  &  51  &  5  \\ 
\hline
\end{tabular}
}
\label{tab:backgroundTableCombo}
\end{table}

The invariant mass distribution of the data is compared 
to the backgrounds and signal templates with pole masses
in the 0.13-2.0~TeV range~\cite{Aad:2011xp, CDF:Zpmumu2fb}. 
A likelihood function is defined as the product of the Poisson
 probabilities over all mass bins in the search region. The Poisson probability in each bin is evaluated for the observed
 number of data events given the background and signal template expectation. The total signal acceptance as a function of mass is propagated into the expectation.

The significance of a signal is summarized by a $p$-value, the probability of observing an excess at least as signal-like as the one observed in data, in the absence of signal. The outcome of the search is ranked using a likelihood ratio,
which is scanned as a function of \zp\ cross section and $m_{\zp}$ over the full considered mass range.
The data are consistent with the SM hypothesis, with $p$-values of 54\% and 24\% for the \ee\ and \mumu\ channels respectively.

 Given the absence of a signal, an upper limit on the signal cross section is determined at the 95\% confidence level (C.L.)
using a Bayesian approach~\cite{bayesianMethod}
 with a flat, positive  prior on the signal cross section.

Mass-dependent systematic uncertainties are incorporated as nuisance parameters which are integrated out~\cite{bayesianMethod}. 
They include normalization to the $Z$-peak, PDF, QCD and weak K-factors, as well as  trigger, reconstruction and identification efficiencies. 
These uncertainties are correlated across all bins in the search region 
and they are correlated between signal and background. 

Since the total background is normalized 
to the data in the region of the $Z\to\ll$  mass peak, the
residual systematic uncertainties are small at 
the $Z$ pole 
and grow at higher mass. The dominant uncertainties are theoretical. 
The overall uncertainty due to PDF and $\alpha_S$ variations is estimated to be 10\% 
at 1.5~TeV using the MSTW 2008 
eigenvector PDF sets and other PDF sets corresponding to variations of $\alpha_S$.
The difference with respect to CTEQ is included as an additional 3\% uncertainty.
The uncertainty on the QCD K-factor is 3\%, evaluated from variations of the renormalization and factorization
scales by factors of two around the nominal values. A systematic uncertainty of 4.5\% is attributed to EW corrections~\cite{Aad:2011xp}.
The uncertainty on the $Z/\gamma^*$ cross section is 5\%, which is applied as a systematic uncertainty on the normalization. 

Experimental systematic effects due to resolution and inefficiencies at high mass were studied. 
In the electron channel, the calorimeter energy resolution is dominated at large $E_{T}$ by a constant term which is 1.2\% in the barrel and 1.8\% in the endcaps, 
with negligible uncertainty. The uncertainty on the resolution in the muon channel is due to residual misalignments and intrinsic position uncertainties in the muon spectrometer that propagate to a change 
in the observed width of the \zp\ (\gstar) line-shape.
The simulation was adjusted to reproduce the data at high muon momentum. The residual uncertainty translates into an event yield uncertainty of less than
1.5\%.
The combined uncertainty on the muon trigger and reconstruction efficiency is estimated to be 4.5\% at 1.5~TeV. 
This uncertainty is dominated by a conservative estimate of the impact of large energy loss from muon bremsstrahlung in the calorimeter on the muon reconstruction performance in the muon spectrometer.
In the electron channel, a systematic uncertainty of 1.5\% at 1.5~TeV is estimated for a possible identification inefficiency caused by the isolation requirement.

\begin{table}[!t]
\caption{Summary of the dominant systematic uncertainties on the expected signal and background yields at $\mll=1.5$~TeV for the \zp\ (\gstar) analysis. 
NA means not applicable. 
}
\label{tab:systematicSummary}
\centering
\addtolength{\tabcolsep}{+1pt}

\resizebox{\columnwidth}{!}{
\begin{tabular}{lcccc}
\hline
Source        & \multicolumn{2}{c}{dielectrons} & \multicolumn{2}{c}{dimuons} \\
	      & signal & background	   & signal & background \\
\hline
Normalization & 5\%		    & NA		      & 5\%	    & NA \\
PDFs/$\alpha_S$ & NA		    & 10\%		      & NA	    & 10\% \\
QCD K-factor  & NA	    & 3\%		      & NA	    & 3\% \\
Weak K-factor & NA		    & 4.5\%		      & NA	    & 4.5\% \\
Trigger/Reconstruction & negligible	    & negligible  & 4.5\%   & 4.5\% \\
\hline
 Total        & 5\%		& 11\%  		 & 7\%  	& 12\%\\
\hline
\end{tabular}
}
\end{table}
The dominant systematic uncertainties are summarized in Table~\ref{tab:systematicSummary}. Uncertainties below
3\% are neglected, and no theory uncertainties are applied to the \zp\ or \gstar\ signal in the limit setting procedure described below. 

The limit on the number of produced \zp\ (\gstar) events is converted into a limit on cross section times branching fraction $\xbr$ 
by scaling with the observed number of \z\ boson events and the theoretical value of $\xbr(Z \rightarrow ll)$.
The expected exclusion limits are determined using simulated pseudo-experiments containing only Standard Model processes,
by evaluating the 95\% C.L. upper limits for each pseudo-experiment for each fixed value of $m_{\zp}$ ($m_{\gstar}$).
The median of the distribution of limits represents the expected limit. The ensemble of limits
is used to find the  68\% and 95\% envelopes of the expected limits as a function of $m_{\zp}$ ($m_{\gstar}$).
Figure~\ref{fig:limit_res} (top) shows the combined dielectron and dimuon 95\% C.L. observed and expected exclusion limits on $\xbr(Z' \rightarrow ll)$.
It also shows the theoretical cross section times branching fraction for the \zpssm\ and for \esix-motivated \zp\ models with the lowest and highest \xbr.
\begin{figure}[!ht]
  \centering
  \includegraphics[width=\columnwidth]{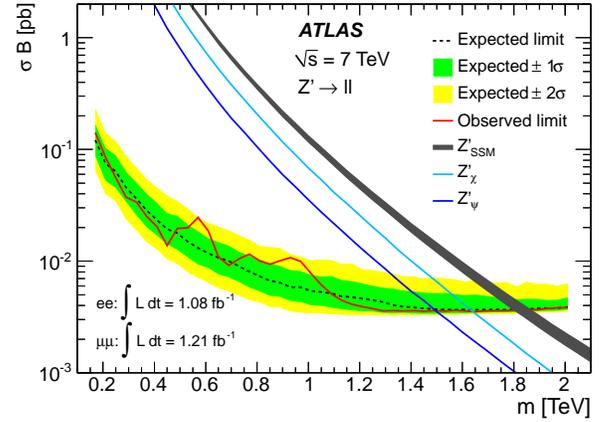}
  \includegraphics[width=\columnwidth]{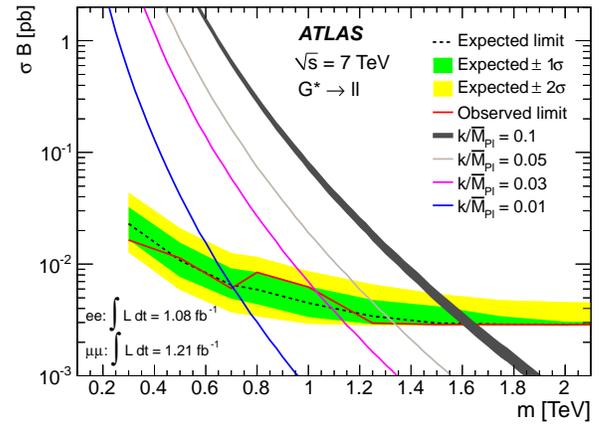}
  \caption{Expected and observed 95\% C.L. upper limits on \xbr\ as a function of mass for \zp\ (top) and $G^*$ (bottom) models. 
           Both results show the combination of the electron and muon channels. 
           The thickness of the \zpssm\ (top) and the \gstar\ for $k/\overline{M}_{Pl}$=0.1 (bottom) theory curves illustrate the theoretical uncertainties.
         }
  \label{fig:limit_res}
\end{figure}
Figure~\ref{fig:limit_res} (bottom) shows the corresponding limits on the RS graviton. 
Mass limits obtained for the \zpssm\ and $G^*$ (with $k/\overline{M}_{Pl}$=0.1) are displayed in Table~\ref{tab:limits}.
The combined mass limits on the \esix-motivated models and the \gstar\ with various couplings are given in Table~\ref{tab:combinedLimits}.
\begin{table}[!htb]
\caption{Observed (Expected) 95\% C.L. mass lower limits in TeV on \zpssm\ resonance and $G^*$ graviton (with $k/\overline{M}_{Pl}$=0.1).
}
\label{tab:limits}
\begin{center}
\small
\begin{tabular}{l c c c}
\hline
Model~~~~~          &      $ \ee$~~~      &       $\mumu$~~~  &   $ \ll$~~   \\
\hline			 	 	 	 	  	 
$\zpssm  $                  &   \LimitElectron~(\LimitElectronExpected)  &    \LimitMuon~(\LimitMuonExpected) &   \LimitCombined~(\LimitCombinedExpected) \\
$G^* $                     &   \LimitElectronG~(\LimitElectronExpectedG)     &    \LimitMuonG~(\LimitMuonExpectedG) 	    	&	\LimitCombinedG~(\LimitCombinedExpectedG)      \\
\hline
\end{tabular}
\end{center}
\end{table}
\begin{table}[!hbt]
\caption{95\% C.L. lower limits on the masses of \esix-motivated \zp\ bosons and RS gravitons \gstar\ for various values of the coupling $k/\overline{M}_{Pl}$.
Both lepton channels are combined.
}
\label{tab:combinedLimits}
\begin{center}
\resizebox{\columnwidth}{!}{
\small
\begin{tabular}{l|cccccc|cccc}
\hline
& \multicolumn{6}{c|}{\esix\ \zp\ Models}   & \multicolumn{4}{c}{RS Graviton} \\
\hline
Model/Coupling            & \zppsi & \zpN  & \zpeta & \zpI  & \zpsq & \zpchi & 0.01 & 0.03  & 0.05 & 0.1 \\
\hline			 	 	 	 	  	 
Mass limit [TeV] & \LimitCombinedPsi  & \LimitCombinedN  & \LimitCombinedEta  & \LimitCombinedI & \LimitCombinedSq & \LimitCombinedChi & \LimitCombinedGOne & \LimitCombinedGThree  & \LimitCombinedGFive  & \LimitCombinedG  \\
\hline
\end{tabular}
}
\end{center}
\end{table}

In conclusion, the ATLAS detector has been used to search for narrow, heavy resonances in the dilepton invariant mass spectrum above the $Z$ boson pole. 
Proton-proton collision data with \LumiElectronsInFb\ (\LumiMuonsInFb)~\ifb\ in the \ee\  (\mumu) channel have been used. 
The observed invariant mass spectra are consistent with the SM expectations. Limits are set on the cross section times branching fraction $\xbr$. 
The resulting mass limits are \LimitCombined~TeV for the Sequential Standard Model \zp\ boson, \LimitCombinedPsi$-$\LimitCombinedChi~TeV for various 
\esix-motivated \zp\ bosons, and \LimitCombinedGOne~TeV$-$\LimitCombinedG~TeV for a Randall-Sundrum graviton with couplings ($k/\overline{M}_{Pl}$) in the range 0.01$-$0.1. 
The \zp\ boson limits are the most stringent to date, including indirect limits set by LEP2.

We thank CERN for the very successful operation of the LHC, as well as the
support staff from our institutions without whom ATLAS could not be
operated efficiently.

We acknowledge the support of ANPCyT, Argentina; YerPhI, Armenia; ARC,
Australia; BMWF, Austria; ANAS, Azerbaijan; SSTC, Belarus; CNPq and FAPESP,
Brazil; NSERC, NRC and CFI, Canada; CERN; CONICYT, Chile; CAS, MOST and
NSFC, China; COLCIENCIAS, Colombia; MSMT CR, MPO CR and VSC CR, Czech
Republic; DNRF, DNSRC and Lundbeck Foundation, Denmark; ARTEMIS, European
Union; IN2P3-CNRS, CEA-DSM/IRFU, France; GNAS, Georgia; BMBF, DFG, HGF, MPG
and AvH Foundation, Germany; GSRT, Greece; ISF, MINERVA, GIF, DIP and
Benoziyo Center, Israel; INFN, Italy; MEXT and JSPS, Japan; CNRST, Morocco;
FOM and NWO, Netherlands; RCN, Norway; MNiSW, Poland; GRICES and FCT,
Portugal; MERYS (MECTS), Romania; MES of Russia and ROSATOM, Russian
Federation; JINR; MSTD, Serbia; MSSR, Slovakia; ARRS and MVZT, Slovenia;
DST/NRF, South Africa; MICINN, Spain; SRC and Wallenberg Foundation,
Sweden; SER, SNSF and Cantons of Bern and Geneva, Switzerland; NSC, Taiwan;
TAEK, Turkey; STFC, the Royal Society and Leverhulme Trust, United Kingdom;
DOE and NSF, United States of America.

The crucial computing support from all WLCG partners is acknowledged
gratefully, in particular from CERN and the ATLAS Tier-1 facilities at
TRIUMF (Canada), NDGF (Denmark, Norway, Sweden), CC-IN2P3 (France),
KIT/GridKA (Germany), INFN-CNAF (Italy), NL-T1 (Netherlands), PIC (Spain),
ASGC (Taiwan), RAL (UK) and BNL (USA) and in the Tier-2 facilities
worldwide.

\begin{widetext}

\clearpage
\begin{flushleft}
{\Large The ATLAS Collaboration}

\bigskip

G.~Aad$^{\rm 48}$,
B.~Abbott$^{\rm 111}$,
J.~Abdallah$^{\rm 11}$,
A.A.~Abdelalim$^{\rm 49}$,
A.~Abdesselam$^{\rm 118}$,
O.~Abdinov$^{\rm 10}$,
B.~Abi$^{\rm 112}$,
M.~Abolins$^{\rm 88}$,
H.~Abramowicz$^{\rm 153}$,
H.~Abreu$^{\rm 115}$
% ...
\bigskip

% ...
$^{10}$ Institute of Physics, Azerbaijan Academy of Sciences, Baku, Azerbaijan\\
$^{11}$ Institut de F\'isica d'Altes Energies and Departament de F\'isica de la Universitat Aut\`onoma  de Barcelona and ICREA, Barcelona, Spain\\
% ...
$^{48}$ Fakult\"{a}t f\"{u}r Mathematik und Physik, Albert-Ludwigs-Universit\"{a}t, Freiburg i.Br., Germany\\
$^{49}$ Section de Physique, Universit\'e de Gen\`eve, Geneva, Switzerland\\
% ...
$^{88}$ Department of Physics and Astronomy, Michigan State University, East Lansing MI, United States of America\\
% ...
$^{111}$ Homer L. Dodge Department of Physics and Astronomy, University of Oklahoma, Norman OK, United States of America\\
$^{112}$ Department of Physics, Oklahoma State University, Stillwater OK, United States of America\\
% ...
$^{115}$ LAL, Univ. Paris-Sud and CNRS/IN2P3, Orsay, France\\
% ...
$^{118}$ Department of Physics, Oxford University, Oxford, United Kingdom\\
% ...
$^{153}$ Raymond and Beverly Sackler School of Physics and Astronomy, Tel Aviv University, Tel Aviv, Israel\\
% ...
\end{flushleft}


\clearpage
\end{widetext}


\begin{thebibliography}{47}
\bibitem[{coo()}]{coordinates}
\bibinfo{note}{ATLAS uses a right-handed coordinate system with the $z$-axis
  along the beam pipe. The $x$-axis points to the center of the LHC ring, and
  the $y$ axis points upward. Cylindrical coordinates $(r,\phi)$ are used in
  the transverse plane, $\phi$ being the azimuthal angle. The pseudorapidity is
  defined in terms of the polar angle $\theta$ as $\eta=-\ln\tan(\theta/2)$.}
\end{thebibliography}
\end{document}